\documentclass[aps,pra,twocolumn,superscriptaddress,longbibliography]{revtex4-2}

\pdfoutput=1
\usepackage[english]{babel}
\usepackage{xcolor}
\newcommand{\comment}[1]{}
\usepackage{amsmath}
\usepackage{amsfonts}
\usepackage{tensor}
\usepackage{physics}
\usepackage{graphicx}
\usepackage[colorlinks=true, allcolors=blue]{hyperref}
\usepackage{siunitx}

\begin{document}

\title{Pulse engineering via projection of response functions}
\author{Nicolas Heimann}
\email{nheimann@physnet.uni-hamburg.de}
\affiliation{Zentrum f\"ur Optische Quantentechnologien, Universit\"at Hamburg, 22761 Hamburg, Germany}
\affiliation{Institut für Quantenphysik, Universit\"at Hamburg, 22761 Hamburg, Germany}
\affiliation{The Hamburg Centre for Ultrafast Imaging, 22761 Hamburg, Germany}
\author{Lukas Broers}
\affiliation{Zentrum f\"ur Optische Quantentechnologien, Universit\"at Hamburg, 22761 Hamburg, Germany}
\affiliation{Institut für Quantenphysik, Universit\"at Hamburg, 22761 Hamburg, Germany}
\author{Ludwig Mathey}
\affiliation{Zentrum f\"ur Optische Quantentechnologien, Universit\"at Hamburg, 22761 Hamburg, Germany}
\affiliation{Institut für Quantenphysik, Universit\"at Hamburg, 22761 Hamburg, Germany}
\affiliation{The Hamburg Centre for Ultrafast Imaging, 22761 Hamburg, Germany}

\begin{abstract}
We present an iterative optimal control method of quantum systems, aimed at an implementation of a desired operation with optimal fidelity. The update step of the method is based on the linear response of the fidelity to the control operators, and its projection onto the mode functions of the corresponding operator. Our method extends methods such as gradient-ascent pulse engineering (GRAPE) and variational quantum algorithms, by determining the fidelity gradient in a hyperparameter-free manner, and using it for a multiparameter update, capitalizing on the multimode overlap of the perturbation and the mode functions.
This directly reduces the number of dynamical trajectories that need to be evaluated in order to update a set of parameters.
We demonstrate this approach, and compare it to the standard GRAPE algorithm, for the example of a quantum gate on two qubits, demonstrating a clear improvement in convergence and optimal fidelity of the generated protocol.
\end{abstract}

\maketitle

\section{Introduction}
\label{sec:introduction}

Algorithmic control and optimized utilization of quantum computational devices plays a central role in the research of quantum technologies.
In recent years, the emergent field of quantum machine learning has brought forth new optimization heuristics that expand the body of quantum optimal control (QOC)~\cite{Peirce88,Brif10,Koch22}. 
One key method is gradient-ascent pulse engineering (GRAPE)~\cite{Palao03,khaneja_optimal_2005}, which is a QOC method that is based on estimating gradients in a space of control parameters to navigate the error surface of a given objective.
As such, GRAPE has found utilization in the context of quantum computing~\cite{rebentrost_optimal_2009,egger_2013_optimizing,jandura_time-optimal_2022,heimann2023quantum}.
Further, variational quantum algorithms (VQAs)~\cite{McClean16,Cerezo2021-Nature-Reviews-Physics} provide a more recent circuit-based approach to QOC~\cite{Li17,Choquette21,Magann21,Meitei21,Keijzer23}  in the context of parametrized quantum circuits.

Gradient-based optimization methods, such as GRAPE, rely on finite differences and large statistics in online utilization on quantum devices, both of which are prone to errors. 
The multitude of dynamic trajectories (or runs) that have to be realized, either in the numerical simulation that is employed or on a real device, are a bottleneck of gradient based heuristics~\cite{bittel2022,Wierichs2022}.
While VQAs are kept in high regard as a promising utilization of noisy intermediate-scale quantum (NISQ) devices~\cite{Preskill2018,Bharti22}, they have been repeatedly shown to display serious shortcomings~\cite{mcclean18,Bittel21,Holmes22,anschuetz22}.
This has highlighted the necessity for extensions of VQA methods~\cite{Grant2019,cerezo21,Holmes22,lbroers_fourier_barren_2024}.

\begin{figure}[t]
    \centering
    \includegraphics[width=7cm]{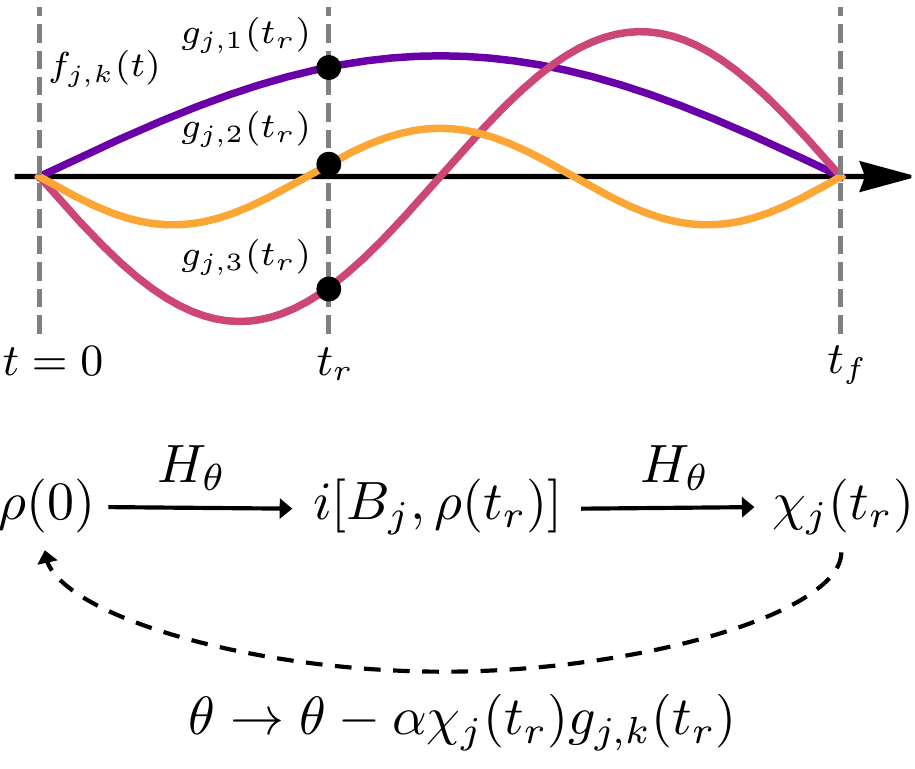}
    \caption{\textbf{Illustration of the PEPR method.}
        Panel~(a) shows temporally nonlocal mode functions $f_{j,k}(t)$ of the control functions that parametrize the time-dependent Hamiltonian.
        The objective is to optimize the fidelity of a desired operation, which is done via iterative updates of the trainable parameters. These updates are based on the response to a perturbation at a random time $t_r$, and the projection of the response onto the mode function $f_{j,k}(t)$, which uses the projection coefficients $g_{j,k}(t_r)$.
        In panel (b) we display a single update schematically. The random initial state $\rho(0)$ is propagated to the random time $t_r$. The commutator with one of the control operators $B_j$ is taken. Then, the resulting density operator is propagated to the time $t_f$, and the response function $\chi_j(t_r)$ is determined. This quantity, together with the projection coefficient $g_{j,k}(t_r)$ and a learning rate $\alpha$, is used for the update.
    }
    \label{fig:1}
\end{figure}
In this paper, we demonstrate a pulse-engineering method based on linear-response functions of time-local perturbations for parametrizations of the Hamiltonian that are nonlocal in time. 
We refer to our method as pulse engineering via projection of response functions (PEPR). The objective of this method is to generate an implementation of a desired operation, such as a quantum algorithm, with optimal fidelity. We assume that 
the quantum system can be controlled via a set of control terms, composed of a set of operators and control functions.
We expand each of these control functions into a set of mode functions, which have an arbitrary time dependence in general, and are temporally nonlocal, in particular. 
In each update step, we determine the response of the fidelity to one of the control operators. This generates the gradient of the fidelity in a hyperparameter-free manner. 
Next, we determine the optimal update of the control parameters via the projection of the response function on the modes of the control functions, resulting in a multiparameter update.
These key features of the method directly address the bottleneck of growing parameter spaces in gradient-based methods, by reducing the necessary amount of dynamical trajectories to iterate over parameter updates. 
Another advantage is the above mentioned elimination of a hyperparameter associated with finite difference methods, improving the usability in a practical setting.

This paper is structured as follows.
In Sec.~\ref{sec:method} we introduce the PEPR method.
In Sec.~\ref{sec:model} we describe the main example to demonstrate the method, which is the optimal implementation of a CNOT gate on a two-qubit system driven by Rabi pulses and coupled via Heisenberg interaction.
In Sec.~\ref{sec:results} we benchmark the optimization performance of PEPR in relation to GRAPE and discuss the optimal solutions obtained in the presence of dissipation and constraints of the system.
In Sec.~\ref{sec:conclusion} we conclude our findings.

\section{Method}
\label{sec:method}

The pulse-engineering method PEPR that we propose can be generally applied to conjugate sets of mode functions of the control functions for the perturbation and the parametrization to maximize the projection coefficients.
Here, we opt for a time-local perturbation and consequently a temporally nonlocal parametrization based on Fourier modes~\cite{lbroers_fourier_barren_2024}.
We consider this approach to be an extension of GRAPE-like methods and VQAs that utilizes nonlocal parametrizations in an efficient manner and addresses common issues, such as gradient estimation and scaling behavior of gradient-based methods. 

We compare the performance of PEPR to that of standard GRAPE in a minimal example of compiling an entangling gate on two qubits. As we discuss below, we find that both the fidelity of the generated protocol, as well as the convergence towards this protocol, is improved in a convincing fashion. We note that QOC has been related to linear-response theory in~\cite{castro11}.

The method works as follows, see Fig.~\ref{fig:1} for an illustration.
Analogously to variational and optimal control approaches, we consider a Hamiltonian of the form
\begin{eqnarray}
H_\theta &=& H_{0} + \sum_{j=1}^{n_{B}} \theta_{j}(t) B_{j}.
\label{eq:hamilton-general}
\end{eqnarray}
The control operators $B_{j}$ represent the options of external control of the system, with $n_{B}$ being the number of them.
We write general parametrizations of the time-dependent control functions $\theta_{j}(t)$ as
\begin{eqnarray}
\theta_{j}(t) &=& \sum_{k=1}^{n_{j}} \theta_{j,k} f_{j,k}(t),\label{thetaj}
\end{eqnarray} 
where $\theta_{j,k}$ are parameters, and $f_{j,k}(t)$ are mode functions of the control functions of the parametrization.
The term $H_{0}$ in Eq.~(\ref{eq:hamilton-general}) is the part of the Hamiltonian that cannot be controlled externally. 
 
We use this Hamiltonian to propagate an initial density matrix $\rho(0)$ over the time interval of $t \in [0, t_{f}]$, such that we obtain the density operator $\rho(t_{f})$. 
The objective of the algorithm is to maximize the fidelity of the state $\rho(t_f)$, compared to the target state that a target transformation $V$ produces, i.e., $\rho^*=V\rho(0)V^\dagger$.
We therefore write the state fidelity as 
\begin{eqnarray}
F_{\theta}(\rho) &=& \Tr(\rho(t_{f})\rho^*),
\label{eq:fidelity-rho}
\end{eqnarray} 
where the subscript $\theta$ emphasizes the dependence on the parameters which determine the time evolution that produces $\rho(t_{f})$.
Equivalently, we aim to minimize the infidelity $1-F_{\theta}(\rho)$ for any initial state $\rho(0)$.
For this purpose, we approach an optimal or near-optimal implementation in an iterative fashion. 

Next, we consider a perturbation of the system of the form
\begin{eqnarray}
H_{p} &=& - \epsilon \delta(t-t_{r}) B_{j}.\label{Hpert}
\end{eqnarray}
The perturbation occurs at time $t_{r}$, with the time dependence of a $\delta$ function. 
$B_{j}$ is one of the control operators via which the system can be controlled as presented in Eq.~(\ref{eq:hamilton-general}).
The prefactor $\epsilon$ is utilized as a perturbative expansion parameter.
We choose the time randomly in the interval $t_r \in [0, t_{f}]$, and we choose the operator $B_{j}$ randomly, via a random choice of the index $j= 1,\ldots, n_{B}$.
 
We then follow the standard result of linear response theory, in which the change of the expectation value of an observable $A$, due to a perturbation of the form $H_{p} = - F(t) B$ is of the general form
\begin{eqnarray}
\Delta \langle  A \rangle &=& \int_{-\infty}^{\infty} dt' \chi_{AB}(t,t')F(t'),\label{DeltaA}
\end{eqnarray}
with the susceptibility
\begin{eqnarray}
\chi_{AB}(t,t') &=& \frac{i}{\hbar} \Theta(t-t') \langle [ A_{I}(t), B_{I}(t')] \rangle,
\end{eqnarray}
where $A_{I}(t)$ and $B_{I}(t')$ are the operators $A$  and $B$ in the interaction picture, respectively. 

We apply this approach to the optimization objective mentioned above. 
In this context, the observable $A$ is the target state $\rho^*$. 
The time at which this observable is evaluated is $t_{f}$, such that we have $A_{I}(t_{f}) = U^{\dagger}(t_{f}) \rho^* U(t_{f})$, where $U(t_{f}) = U(0, t_{f})$ is the time-evolution operator from time $0$ to time $t_{f}$.
The perturbation contains one of the control operators $B_{j}$, and $F(t) = \epsilon \delta(t-t_{r})$.
We therefore have $B_{j,I}(t_{r}) = U^{\dagger}(t_{r}) B_j U(t_{r})$.
With this, we write
\onecolumngrid\begin{eqnarray}
\chi_{j}(t_{r}) \equiv \chi_{F_{\theta}, B_{j}}(t_{f}, t_{r}) &=& \frac{i}{\hbar} \Tr\Big( \Big[  U^{\dagger}(t_{f}) \rho^* U(t_{f}), U^{\dagger}(t_{r}) B_j U(t_{r})\Big] \rho(0) \Big)\\
&=& \frac{i}{\hbar} \Tr\Big( \rho^*  U(t_{r}, t_{f}) \Big[  B_{j},  U(t_{r}) \rho(0) U^{\dagger}(t_{r})\Big]  U^{\dagger}(t_{r}, t_{f}) \Big),\label{eq:chi-final}
\end{eqnarray}\twocolumngrid
where we note that $t_{f}>t_{r}$, such that $\Theta(t_{f}-t_{r})=1$.
Furthermore, using Eq.~(\ref{DeltaA}) and $F(t) = \epsilon \delta(t-t_{r})$, we obtain
\begin{eqnarray}
\frac{\Delta  F_{\theta}}{\epsilon} &=& \chi_{j}(t_{r}).
\end{eqnarray}
This expression shows that the gradient of the fidelity in the operator space, spanned by the control operators $B_{j}$, is determined by computing the linear response of the fidelity with regard to that operator at a time $t_{r}$.
  
Based on this response function, we determine the optimal update of the parameters $\theta_{j,k}$ as follows, where the index $j$ corresponds to the operator $B_{j}$ in Eq.~(\ref{Hpert}).
We write the projection of the $\delta$ function that is used in the perturbation in Eq.~(\ref{Hpert}) onto the mode functions of the control functions of the parametrization as
\begin{eqnarray}
\delta(t-t_{r}) &=& \sum_{k=1}^{\infty} g_{j,k}(t_{r}) f_{j,k} (t).\label{deltadecomp}
\end{eqnarray}
The $g_{j,k}(t)$ are conjugate functions to the $f_{j,k}(t)$ in the sense of a decomposition of the $\delta$ function.
Based on this decomposition of the time dependence of the perturbation in terms of the mode functions of the control function of the control operator $B_j$, we obtain the update rule
\begin{eqnarray}
\theta_{j,k} &\rightarrow & \theta_{j,k} - \alpha_0 g_{j,k}(t_{r}) \chi_{j}(t_{r}),
\label{eq:theta-update-lr}
\end{eqnarray}
where we introduce the parameter $\alpha_0$, which in similar contexts is referred to as a learning rate or step size.
While the decomposition in Eq.~(\ref{deltadecomp}) is exact, we use the approximation of truncating the sum by the number of modes $n_{j}$ that are included in the representation in Eq.~(\ref{thetaj}). 
With this, we identify the term $- \alpha_0 g_{j,k}(t_r) \chi_j(t_r)$ as a correction to the parameters $\theta_{j,k}$.
We emphasize that $\chi_j(t_r)$ in Eq.~(\ref{eq:theta-update-lr}) is used to update $n_{j}$-many parameters $\theta_{j,k}$ without increasing the numerical complexity of the approach, assuming that the corresponding $g_{j,k}(t_r)\not=0$.
This is in contrast to conventional variational methods, in which the complexity grows with the controllability of the Hamiltonian.
Therefore, this method directly benefits from parametrizations in which the mode functions of the control functions have significant overlap with the time dependence of the perturbation, i.e., the $\delta$ functions acting at different times.

Note that $\theta_j(t)$ are functions that can be arbitrarily parametrized through choices of $f_{j,k}(t)$ in Eq.~(\ref{thetaj}).
Common parametrizations in optimal control contexts use time-local stepwise functions, which is adjacent to parametrized variational quantum circuit methods, or low-dimensional random bases~\cite{Doria11,Caneva11}. 
Since this method is an extension of GRAPE, and relies on, and benefits from, the overlap between different functional bases for the perturbation and the parametrization of the Hamiltonian, we refer to it as pulse engineering via the projection of response functions (PEPR).
Hence, a central aspect of PEPR is the overlap of parametrization mode functions of the control functions and the perturbation-mode functions of the control functions.

As an example that implements these considerations, we choose the temporally nonlocal parametrization of Fourier modes~\cite{lbroers_fourier_barren_2024}, which consists of mode functions of the control functions that all have overlap with (almost) any time-local perturbation proportional to $\delta(t-t')$.
This particular parametrization is
\begin{equation}
\theta_j(t)=\sum_{k=1}^{n_{j}}\theta_{j,k}\sin(\pi k \frac{t}{t_\mathrm{f}}).
\label{eq:theta-sine}
\end{equation}
This means that the decomposition of the $\delta$ function in Eq.~(\ref{deltadecomp}) is now given through the functions
\begin{eqnarray}
  f_{j,k}(t) &=& \sin(\pi k \frac{t}{t_\mathrm{f}}),\\
  g_{j,k}(t) &=& \frac{2}{t_\mathrm{f}}\sin(\pi k \frac{t}{t_\mathrm{f}}),
\end{eqnarray}
such that we obtain the update rule
\begin{eqnarray}
  \theta_{j,k} &\rightarrow & \theta_{j,k} - \alpha \sin(\pi k \frac{t_{r}}{t_\mathrm{f}}) \chi_{j}(t_{r}),\label{eq:theta-update-lr-sine}
\end{eqnarray}
where we have introduced the effective learning rate $\alpha~=~\alpha_0 \frac{2}{t_\mathrm{f}}$. 

\section{Model}
\label{sec:model}

\begin{figure}[t]
    \centering
    \includegraphics{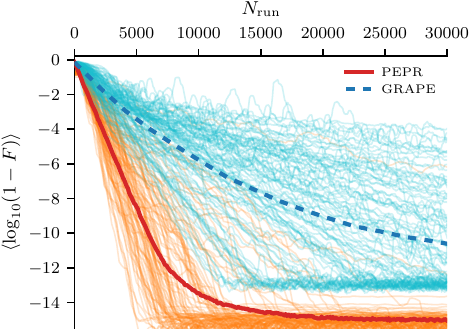}
    \caption{\textbf{PEPR vs GRAPE.}
        The average infidelity $\langle \text{log}_{10}(1-F)\rangle$, as a function of the number of runs $N_\text{run}$, obtained using PEPR (red line) and GRAPE (blue dashed).
        The underlying ensembles of realizations $\{1-F_\theta\}$ for PEPR and GRAPE are depicted as orange and cyan lines, respectively.
        PEPR leads to faster convergence with a reduced variance of the underlying realizations, compared to GRAPE.
        Furthermore, the infidelity that is achieved is consistently smaller for PEPR than for GRAPE.
        The lower bound of the average infidelity is determined by the accuracy of the used numerical method.
        Due to the additional error of the gradient estimate, this is more detrimental for GRAPE.}
    \label{fig:2}
\end{figure}

We demonstrate PEPR for a simple example of the target transformation $V=\text{CNOT}$ on two qubits.
The full Hamiltonian reads
\begin{equation}
    H(t) = \sum_{j=1}^2 h_{x,j}(t)\sigma_x^j + h_{y,j}(t) \sigma_y^j + J(t)\vec{\sigma}^1\vec{\sigma}^2,
\end{equation}
where $\vec{\sigma}$ is the vector of Pauli matrices.
The time-dependent functions are equivalent to the $\theta_j(t)$ in Eq.~(\ref{thetaj}).
We parametrize these functions as described in Eq.~(\ref{eq:theta-sine}) such that 
\begin{eqnarray}
    h_{x/y, j}(t)&=&\sum_{k=1}^{n_{j}} \theta_{x/y,j,k}\sin(\pi k \frac{t}{t_f}),\\
    J(t)&=&\sum_{k=1}^{n_{j}} \theta_{J,k}\sin(\pi k \frac{t}{t_f}).
\end{eqnarray}

Next we describe two generic scenarios for optimization tasks, namely, optimization with and without constraints on the parameters.

For optimization without constraints, we begin by randomly sampling the initial parameters $\theta$ from normal distributions, i.e., $\theta_{j,k}\sim\mathcal{N}(0,1)$.
We then follow the PEPR procedure, described above, to iteratively update these parameters in order to identify parameters that produce a time evolution that implements the target transformation $V$.
In each iteration, we initialize the state of the system in a product state $\rho(0)=\rho_1\otimes\rho_2$, where $\rho_1$ and $\rho_2$ are random local pure states, see App.~\ref{app:CNOT} for details.
We randomly choose the control operator $B_j\in\{\sigma_{x}^{1},\sigma_{y}^{1},\sigma_{x}^{2}, \sigma_{y}^{2}, \vec{\sigma}^1\vec{\sigma}^2\}$ and evaluate the susceptibility $\chi_j(t_r)$ for a random time $t_r\in [0,t_f]$.
We then update the parameters according to Eq.~(\ref{eq:theta-update-lr-sine}).

For optimization with constraints, we discuss the example that the amplitude of the Rabi pulses is smaller than an upper bound $\Omega_\text{max}$, so we demand $|h_{x,j}(t) - i h_{y,j}(t)| < \Omega_\text{max}$.
Similarly, we demand that the magnitude of the Heisenberg coupling $J(t)$ is smaller than an upper bound $J_\text{max}$, namely, we demand $|J(t)|<J_\text{max}$.
We implement the constraints for PEPR and GRAPE as follows.

In an update step of PEPR, we generate the potential update of the parameters, and then check if the new set of parameters violates the constraints or not.
If they do, we discard the potential update, and generate a new potential update.
If the new set of parameters fulfills the constraints, the parameter update is accepted.
To generate initial values for the parameters that fulfill the constraints, we first sample the parameters $\theta$ from normal distributions, i.e. $\theta_{x/y, j, k}\sim\mathcal{N}(0,1)$ and $\theta_{J, k}\sim\mathcal{N}(0,1)$.
We then check if these potential initial values fulfill the constraints.
If they do not, we rescale the parameters via $\theta_{x/y, j, k} \rightarrow \theta_{x/y, j, k}/\text{max}(1, \text{max}_t (|h_{x,j}(t) - i h_{y,j}(t)|)/\Omega_\text{max})$, and $\theta_{J,k} \rightarrow \theta_{J, k} / \text{max}(1,\text{max}_t (|J(t)|)/J_\text{max})$.

For an update step of GRAPE, we rescale the control functions via $h_{x/y,j}(t)\rightarrow h_{x/y,j}(t)/\max(1, |h_{x,j}(t)- i h_{y,j}(t)|/\Omega_\text{max})$ and $J(t)\rightarrow J(t)/\max(1, |J(t)|/J_\text{max})$.
This truncates the maximal value of $|h_{x,j}(t) - i h_{y,j}(t)|$ and $|J(t)|$ at the values of $\Omega_\text{max}$ and $J_\text{max}$, respectively.
To generate initial values for the parameters theta for the GRAPE method, we sample them randomly via $\theta_{x/y, j, k}\sim\mathcal{N}(0,1)$ and $\theta_{J,k}\sim\mathcal{N}(0,1)$.
Then, we rescale them in the same way as for the PEPR algorithm.

The difference of implementing the constraints in the GRAPE method derives from PEPR being a stochastic method, such that a rejection of a potential update does not terminate the algorithm, while GRAPE is a deterministic method, in which a rejection terminates the algorithm.
Thus, we implement constraints in GRAPE differently, as described.

Up to this point, we have described a method that operates using a state fidelity.
By averaging over these fidelities we obtain a metric that is representative of the fidelity of the unitary transformation realized through the time evolution.
We therefore approximate the infidelity of the transformation by considering $n_F$ trajectories $\{\rho\}$, obtained from different initial states $\rho(0)$, and averaging over their state fidelities, such that
\begin{eqnarray}
    1-F_\theta =1-\frac{1}{n_F}\sum_{\{\rho\}}F_\theta(\rho).
\end{eqnarray}
$F_\theta(\rho)$ is the state fidelity according to Eq.~(\ref{eq:fidelity-rho}), where the subscript emphasizes the dependence on the parameters $\theta$. 
Here we choose an empirical sampling size of $n_F=10$.
The result of this method, as well as similar optimization approaches, is intrinsically probabilistic. 
In particular, the random initial parameters $\theta$ are the starting point for the local optimization sequence within the parameter space. 
In order to present a meaningful comparison, we show many realizations of this algorithm with $n_T$ different initial parameters $\{\theta\}$.
We also show the average infidelity over this ensemble of optimization trajectories. We write
\begin{eqnarray}
    \langle \log_{10}(1-F) \rangle = \frac{1}{n_T}\sum_{\{\theta\}} \log_{10}(1-F_\theta).
    \label{eq:infidelity-average}
\end{eqnarray}
Here we choose an empirical sampling size of $n_T=100$.
Note that this is the log-mean of the infidelity, which gives higher weight to low-infidelity realizations.
We choose this more-involved average as visual support for the set of trajectories in Fig.~\ref{fig:2}, see below.

\begin{figure}[t]
    \centering
    \includegraphics{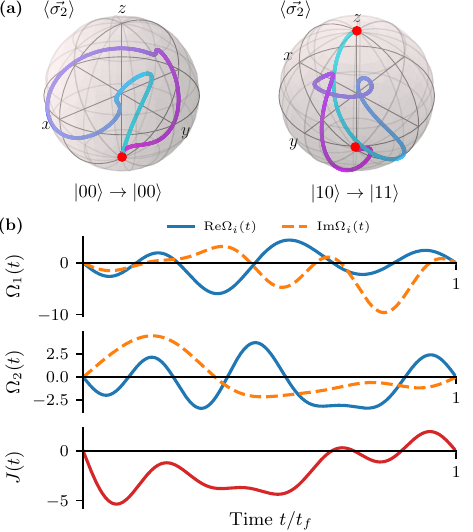}
    \caption{\textbf{PEPR-based high-fidelity protocol.}
        In panel (a) we show the time evolution of the Bloch vector of the target qubit $\langle\vec{\sigma}\rangle$ under a high-fidelity implementation of the CNOT transformation for $\ket{00}\rightarrow\ket{00}$ and $\ket{10}\rightarrow\ket{11}$.
        The time $t$ is depicted as the line color and the initial and final states are depicted as red dots.
        In panel (b) we show the corresponding control functions $\Omega_i(t)=h_{x,i}(t)-i h_{y,i}(t)$ and $J(t)$.
        We note that the amplitudes $|\Omega_i(t)|$ and $J(t)$ are within reasonable boundaries for the given energy scales of the system, defined by the final time $t_f$.}
    \label{fig:3}
\end{figure}

\section{Results}
\label{sec:results}

In this section we demonstrate our method in the case of the target transformation of the CNOT gate.
As a comparative benchmark, we additionally optimize the parameters based on the GRAPE method, see App.~\ref{app:grape} for details.
In the version of GRAPE used here, we update all the parameters in a single iteration.
Therefore, GRAPE requires calculating $5n_j+1$ time evolutions to estimate the gradient of the infidelity in a single iteration.
We emphasize this with regards to our definition of the number of runs $N_\text{run}$, which is the number of calculated time evolutions during a sequence of parameter updates.

We first give an example for optimization without constraints.
In Fig.~\ref{fig:2} we show results of the optimization using both PEPR and GRAPE.
We use an optimized set of hyperparameters $\alpha_\text{PEPR}=0.5$, $\alpha_\text{GRAPE}=1.2$, and $\epsilon=10^{-7}$ for each method to provide an unbiased comparison, as discussed in the supplemental material.
The optimization trajectories generated using PEPR converge faster to low values of the infidelity and show a reduced variance for a given number of runs $N_\text{run}$, compared to GRAPE.
The lower bound is determined by the accuracy of the numerical integration method.
Here, we use standard fourth-order Runge-Kutta with a time discretization of $h=2^{-14}t_f$.
The lower bound of the optimization trajectories using GRAPE additionally depends on the finite difference length $\epsilon$.
Note that the standard GRAPE results show a large variance in the quality of optimization trajectories, compared to PEPR, where we more reliably find fast-converging high-fidelity solutions.
The average infidelity over the ensemble of optimization trajectories $\langle \log_{10}(1-F) \rangle$ also reflects this convergence behavior.
We note that the variance $\text{Var}(1-F)$ obtained using PEPR is reduced, compared to GRAPE, as we show in App.~\ref{app:var}.

We show an example for a high-fidelity implementation of the CNOT gate generated via PEPR in Fig.~\ref{fig:3}.
We show the time evolution of the Bloch vector of the target qubit $\langle \vec{\sigma}^2(t)\rangle = (\langle \sigma_x^2(t)\rangle , \langle \sigma_y^2(t)\rangle, \langle  \sigma_z^2(t)\rangle)^T$ for the example of the initial states $\ket{00}$ and $\ket{10}$.
The trajectory of $\langle \vec{\sigma}^2(t)\rangle$ is continuous and efficient on the time scale of $t_f$.
We note that the amplitudes of the control functions are within reasonable boundaries, resulting in a realistic protocol.

\begin{figure}[t]
    \centering
    \includegraphics{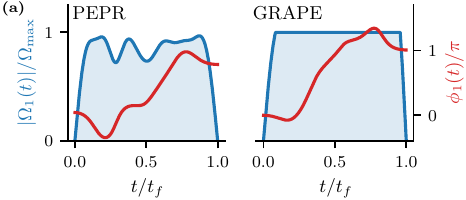}
    \includegraphics{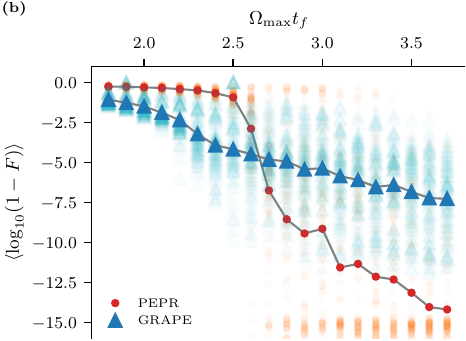}
    \caption{\textbf{Optimization under constraints.}
    In panel~(a) we show high-fidelity Rabi protocols of the first qubit $\Omega_1(t)=|\Omega_1(t)|e^{i\phi_1(t)}$ obtained using PEPR and GRAPE
    optimization under constraints.
    Specifically, we use $\Omega_\text{max}t_f=J_\text{max}t_f=2.7$, and $N_\text{run}~=~\num{30000}$.
    In both cases, we obtain protocols close to phase-only protocols as the corresponding pulse areas are close to the maximal value of $\Omega_\text{max}t_f$.
    In panel~(b) we show the average infidelity $\langle \text{log}_{10}(1-F) \rangle$ as a function of the available pulse area $t_f\Omega_\text{max}$ and with a maximal interaction strength of $J_\text{max}=\Omega_\text{max}$, for $N_\text{run}~=~\num{30000}$, obtained using PEPR (red dots) and GRAPE (blue triangles).
    The underlying ensembles of realizations $\{1 - F_\theta(N_\text{run})\}$ obtained using PEPR and GRAPE are depicted as orange dots and cyan rectangles, respectively.}
    \label{fig:4}
\end{figure}

As a second example, we demonstrate optimization under constraints.
We use the constraint that we discussed above, i.e., $|h_{x,j}(t) - i h_{y, j}(t)| < \Omega_\text{max}$ and $|J(t)|<J_\text{max}$.
In Fig.~\ref{fig:4}~(a) we show the control functions $\Omega_1(t)=h_{x,1}(t)-i h_{y,1}(t)$ of high-fidelity protocols, obtained using PEPR and GRAPE, for $N_\text{run}~=~\num{30000}$ runs, and for $\Omega_\text{max} t_f = J_\text{max} t_f = 2.7$.
The infidelity for the PEPR method is $7.64\times 10^{-7}$, and for GRAPE it is $1.12\times 10^{-4}$, for this example.
In both cases, the pulse area $\Theta_j=\int_0^{t_f} |\Omega_j(t)|dt$ is close to the maximal value of $\Omega_\text{max} t_f$, therefore the protocols approximate phase-only protocols, i.e., $\Omega_1(t)\approx \Omega_\text{max}e^{i\phi_1(t)}$.

To elaborate on the properties of protocols under constraints, we show the infidelities obtained after $N_\text{run}~=~\num{30000}$, in Fig.~\ref{fig:4}~(b), and the average infidelity, as a function of $\Omega_\text{max} t_f$.
For the constraints we choose $\Omega_\text{max} = J_\text{max}$.
We find that for PEPR, the infidelities are high up to around $\Omega_\text{max} t_f \approx 2.6$.
For larger values, the infidelities fall of quickly to small values of $10^{-15}$, constrained primarily by the numerical accuracy of the ODE solver.
In comparison, we find that GRAPE has lower average infidelities for $\Omega_\text{max} t_f < 2.6$, while showing a large variance of the obtained infidelities, but has larger infidelities for $\Omega_\text{max} t_f > 2.6$.
This demonstrates that PEPR has an intrinsically better performance in finding high-quality implementations.

\begin{figure}[t]
    \centering
    \includegraphics{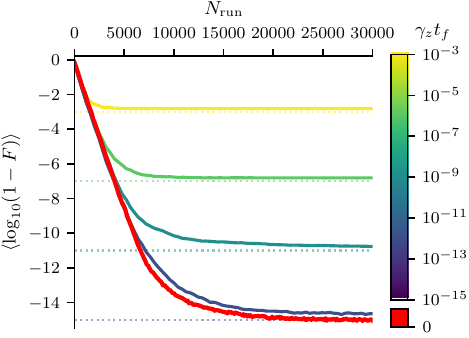}
    \caption{\textbf{Optimization for a system with dissipation.}
        The average infidelity $\langle \text{log}_{10}(1-F) \rangle$, as a function of the number of runs $N_\text{run}$, obtained via PEPR, for different values of the dissipation rate $\gamma_z$.
        In the initial stage of optimization, the average infidelity is equal to the average infidelity of the non-dissipative case (red), which reproduces the data shown in Fig.~\ref{fig:2}.
        In the later stage of optimization, the average infidelity converges to values close to the estimated lower bound of $\gamma_z t_f$ (dotted).
        }
    \label{fig:5}
\end{figure}

To expand the scope of the method, we apply it to a system with dissipation. As a concrete example, we include dephasing with the dissipation rate $\gamma_z$, via the Lindblad master equation
\begin{equation}
    \frac{\partial \rho}{\partial t} = -\frac{i}{\hbar}[H_\theta(t), \rho] + \gamma_z\sum_i \mathcal{D}[\sigma_z^i]\rho.
    \label{eq:meq}
\end{equation}
We apply the PEPR method to this system, as described above.
In Fig.~\ref{fig:5} we show the average infidelity as a function of the number of runs $N_\text{run}$ for different values of the dissipation rate $\gamma_z$.
For comparison, we reproduce the optimization process for vanishing dissipation from Fig.~\ref{fig:2}, displayed in red.
In all cases, the average infidelity $\langle \log_{10}(1-F) \rangle$ converges for $N_\text{run} \gtrsim 2\times 10^4$ runs.
We observe that the dissipation rate introduces a lower bound on the infidelity, as indicated by the dashed lines, which are $\gamma_z t_f$.
However, we find that the method generates optimal protocols in the presence of dissipation, in addition to the dissipationless case discussed above.

\section{Conclusion}
\label{sec:conclusion}

We have presented a quantum optimization method based on linear-response theory, and further based on projecting the response onto control parameters.
Maximizing these projection coefficients results in a more efficient update rule that drastically reduces the number of dynamical realizations of the time evolution necessary to update a set of parameters.
Due to the nature of our approach, we refer to it as pulse engineering via the projection of response functions (PEPR).
We understand this method in the context of the current resurgence of quantum optimal control theory under the moniker of quantum machine learning. 
PEPR is adjacent to GRAPE and VQAs, but extends them efficiently to time-nonlocal parametrizations that directly benefit from the projection feature of the method.
In the endeavor to maximize the overlap of perturbation and parametrization bases, we consider a parametrization that utilizes a low-frequency Fourier-mode expansion.

In a minimal proof-of-concept example we demonstrate the benefits of PEPR over standard GRAPE.
In a direct comparison of optimizing the CNOT gate, we find that PEPR produces high-fidelity solutions more reliably. 
Importantly, PEPR is more efficient in its optimization as it consistently requires less evaluations of time evolution trajectories in order to optimize its protocols.
The evaluation of such time evolutions in order to navigate the parameter space is an important bottleneck of gradient-based methods.
Furthermore, we demonstrate that PEPR is efficient in optimizing systems with dissipation, as well as for optimization under constraints.
Therefore, this improvement of quantum optimal control methods will support the design and establishment of quantum technology going forward.

\begin{acknowledgments}
This work is funded by the Deutsche Forschungsgemeinschaft (DFG, German Research Foundation) - SFB-925 - Project No. 170620586 and the Cluster of Excellence ``Advanced Imaging of Matter'' (EXC 2056) Project No. 390715994.
The project is co-financed by ERDF of the European Union and by ``Fonds of the Hamburg Ministry of Science, Research, Equalities and Districts (BWFGB)''.
\end{acknowledgments} 

\bibliography{main.bib}

\onecolumngrid

\appendix

\section{Gradient Ascent Pulse Engineering}
\label{app:grape}

The objective of GRAPE~\cite{khaneja_optimal_2005} is to minimize a certain loss function defined for a quantum system by inferring an optimal set of parameters $\theta_\text{opt}$ that define the dynamics of that system.
Here, we consider the Hamiltonian $H_\theta$ in Eq.~(\ref{eq:hamilton-general}), which produces the formal time-evolution operator 
\begin{equation}
    U_\theta(0,t_f)=e^{-\frac{i}{\hbar}\int_0^{t_f} H_\theta(t) dt }.
\end{equation}
There are various ways to construct a loss function to evaluate the dynamics with.
For instance, given a target transformation $V$, we can define the state-infidelity
\begin{equation}
    1-F = \text{Tr}(V\rho_0 V^\dagger \rho(t_f)),
\end{equation}
where $\rho_0$ is some initial state and $\rho(t_f)=U(0,t_f)\rho_0U^\dagger(0,t_f)$ is the state obtained from time-propagating the initial state $\rho_0$ from the initial time $t=0$ to $t_f$. 
The central idea behind GRAPE is to update the parameters $\theta$ by gradient ascent (or descent) with respect to the loss function $\mathcal{L}$.
It is possible to approximate the gradient of the loss function $\nabla \mathcal{L}$ by the finite differences
\begin{equation}
    \frac{\partial \mathcal{L}}{\partial \theta_{j,k}} = \frac{\mathcal{L}(\theta+e_{kj}\epsilon)-\mathcal{L}(\theta)}{\epsilon},
    \label{eq:theta-update-grape}
\end{equation}
where $e_{j,k}$ is the unit-vector corresponding to the parameter $\theta_{j,k}$.
This allows us to update the parameters as
\begin{equation}
    \theta \rightarrow \theta - \alpha \nabla\mathcal{L}.
\end{equation}
$\alpha$ is a hyperparameter commonly reffered to as the learning rate in gradient descent contexts.
We note that rather than utilizing the full gradient, the parameters $\theta_{j,k}$ can also be updated individually. 
This requires two evaluation of the loss function per parameter, i.e. $\mathcal{L}(\theta_{j,k}+\epsilon)$ and $\mathcal{L}(\theta_{j,k})$.
Therefore, it is common to update all the parameters within a single iteration of the update rule.
This reduces the computational costs from $2|\theta|$ to $|\theta|+1$ time evolutions, where $|\theta|$ is the number of parameters.

\section{CNOT Gate}
\label{app:CNOT}

In this section we describe the model we use in the main-text to demonstrate the performance of PEPR to obtain high-fidelity realizations of the CNOT gate.
The qubits are locally controlled as
\begin{align}
    H_{1}(t) &= h_{x, 1}(t) \sigma_{x}^{1} + h_{y, 1}(t) \sigma_{y}^{1},\\
    H_{2}(t) &= h_{x, 2}(t) \sigma_{x}^{2} + h_{y, 2}(t) \sigma_{y}^{2},
\end{align}
where $\sigma_{j}^1=\sigma_{j} \otimes 1$, and $\sigma_{j}^2=1\otimes\sigma_{j}$ are the local Pauli matrices.
The interaction-term between the qubits is
\begin{equation}
    H_\text{int}= J(t) \vec{\sigma}_{1} \vec{\sigma}_{2},
\end{equation}
where $ \vec{\sigma}_{1} \vec{\sigma}_{2}=\sigma_x\otimes\sigma_x+\sigma_y\otimes\sigma_y+\sigma_z\otimes\sigma_z$, and $J(t)$ is the time-dependent strength of the interaction.
We write the resulting Hamiltonian as
\begin{equation}
    H_\theta = H_{1} + H_{2} + H_\text{int} = \begin{pmatrix} J  & h_{x, 2}  - i h_{y, 2} & h_{x, 1}  - i h_{y, 1}  & 0\\
 h_{x, 2}  + i h_{y, 2} & -J & 2 J & h_{x, 1}  - i h_{y, 1}\\  
 h_{x, 1}  + i h_{y, 1} & 2J & - J & h_{x, 2}  - i h_{y, 2}\\   
 0  & h_{x, 1}  + i h_{y, 1} & h_{x, 2}  + i h_{y, 2}  & J\\
  \end{pmatrix}.
  \label{eq:hamilton-cnot-app}
\end{equation}
As outlined in the main-text, we consider the parametrized protocols
\begin{align}
    h_{p,i}(t) &= \sum_k \theta_{p,i,k}\sin(\pi k \frac{t}{t_f}), \\
    J(t) &= \sum_k \theta_{J,k}\sin(\pi k \frac{t}{t_f}),
\end{align}
where $p\in\{x,y\}$, $i\in\{1,2\}$.
We explicitly write the density operator of the system as
\begin{equation}
    \rho=\begin{pmatrix} \rho_{1}  & \rho_{4} - i \rho_{5} & \rho_{6} - i \rho_{7} & \rho_{10} - i \rho_{11}\\
  \rho_{4} + i \rho_{5} & \rho_{2}& \rho_{8} - i \rho_{9} & \rho_{12} - i \rho_{13}\\  
 \rho_{6} + i \rho_{7} & \rho_{8} + i \rho_{9} & \rho_{3} & \rho_{14} - i \rho_{15}\\   
  \rho_{10} + i \rho_{11}  & \rho_{12} + i \rho_{13} & \rho_{14} + i \rho_{15}  & \xi-\rho_{1} -\rho_{2} - \rho_{3}
  \label{eq:rho_mat}
  \end{pmatrix}.
\end{equation}
where $\rho_i \in \mathbb{R}$, and $\xi\in\{0,1\}$ defines the value of trace of the state.
We note that $\xi=0$ is required to time-propagate $i[B_j,\rho]$, as $\text{Tr}([B_j,\rho])=0$.
For numerical purpose, we choose to represent $\rho$ as a real-valued vector
\begin{equation}
    \vec{\rho} = (\rho_1, ..., \rho_{15})^T.
\end{equation}
We include dissipation in the form of pure-dephasing of the individual qubits.
The time evolution of the state is governed by the Lindblad master equation
\begin{equation}
    \frac{\partial \rho}{\partial t} = -\frac{i}{\hbar}[H_\theta(t), \rho] + \gamma_z\sum_i \mathcal{D}[\sigma_z^i]\rho,
\end{equation}
where $\mathcal{D}[L]\rho=L\rho L^\dagger-\frac{1}{2}\{L^{\dagger} L, \rho\}$.
The equations of motion read
\begin{align}
 \partial_{t} \rho_{1} =& 2 (- h_{y,2} \rho_{4} + h_{x,2} \rho_{5} - h_{y,1} \rho_{6} + h_{x,1} \rho_{7}),\\
 \partial_{t} \rho_{2} =& 2 (- h_{y,1} \rho_{12} + h_{x,1} \rho_{13} + h_{y,2} \rho_{4} - h_{x,2} \rho_{5} + 2 J \rho_{9}),\\
 \partial_{t} \rho_{3} =& -2 (h_{y,2} \rho_{14} - h_{x,2} \rho_{15} - h_{y,1} \rho_{6} + h_{x,1} \rho_{7} + 2 J \rho_{9}),\\  
\partial_{t} \rho_{4} =& h_{x,1} \rho_{11} + h_{y,2} (\rho_{1} - \rho_{2}) - 2 \gamma_{z} \rho_{4} - 2 J \rho_{5} 
 + 2 J \rho_{7} - h_{y,1} (\rho_{10} + \rho_{8}) + h_{x,1} \rho_{9},\\
 \partial_{t} \rho_{5} =& - h_{y,1} \rho_{11} + h_{x,2}(-\rho_{1} +\rho_{2}) + 2 J \rho_{4} 
  - 2 \gamma_{z} \rho_{5} - 2 J \rho_{6} + h_{x,1}(-\rho_{10} + \rho_{8} ) + h_{y,1} \rho_{9},\\
  \partial_{t} \rho_{6} =& h_{x,2} \rho_{11} + h_{y,1} (\rho_{1} - \rho_{3}) + 2 J \rho_{5} - 2 \gamma_{z} \rho_{6}
   - 2 J \rho_{7} - h_{y,2} (\rho_{10} + \rho_{8}) - h_{x,2} \rho_{9},\\
 \partial_{t} \rho_{7} =& h_{x,1}(-\rho_{1} + \rho_{3}) - 2( J \rho_{4} - J \rho_{6} + \gamma_{z} \rho_{7}) 
 + h_{x,2}(-\rho_{10} + \rho_{8}) - h_{y,2}(\rho_{11} + \rho_{9}),\\
 \partial_{t} \rho_{8} =& - h_{y,1} \rho_{14} + h_{x,1} \rho_{15} + h_{y,1} \rho_{4} - h_{x,1} \rho_{5} + h_{y,2}(-\rho_{12} + \rho_{6}) +h_{x,2} (\rho_{13} - \rho_{7})
  - 4 \gamma_{z} \rho_{8},\\
  \partial_{t} \rho_{9} =& h_{x,1} \rho_{14} + h_{y,1} \rho_{15} - 2 J \rho_{2} + 2 J \rho_{3} - h_{x,1} \rho_{4} - h_{y,1} \rho_{5} + h_{x,2} (-\rho_{12} + \rho_{6})
  + h_{y,2} (-\rho_{13} + \rho_{7}) - 4 \gamma_{z} \rho_{9},\\
 \partial_{t} \rho_{10} =& - 4 \gamma_{z} \rho_{10} - h_{x,2} \rho_{13} - h_{y,1} \rho_{14} - h_{x,1} \rho_{15}
  + h_{y,1} \rho_{4} + h_{x,1} \rho_{5} + h_{y,2}(-\rho_{12} + \rho_{6} )
   + h_{x,2} \rho_{7},\\
 \partial_{t} \rho_{11} =&   - 4 \gamma_{z} \rho_{11} - h_{y,2} \rho_{13} + h_{x,1} \rho_{14} - h_{y,1} \rho_{15} + h_{x,1} \rho_{4} + h_{y,1} \rho_{5} + h_{x,2} (\rho_{12} - \rho_{6}) 
 + h_{y,2} \rho_{7},\\
 \partial_{t} \rho_{12} =&    - h_{x,2} \rho_{11} - 2 \gamma_{z} \rho_{12} + 2 J \rho_{13} - 2 J \rho_{15}
  + h_{y,1} (-1 + \rho_{1} + \rho_{2} + \rho_{3}) + h_{y,2} (\rho_{10} + \rho_{8})
   + h_{x,2} \rho_{9},\\
 \partial_{t} \rho_{13} =&  h_{y,2} \rho_{11} - 2 J \rho_{12} - 2 \gamma_{z} \rho_{13} + 2 J \rho_{14} - h_{x,1}
 (-1 + \rho_{1} + \rho_{2} + \rho_{3}) + h_{x,2} (\rho_{10} - \rho_{8})
  + h_{y,2} \rho_{9},\\ 
  \partial_{t} \rho_{14} =&   - 2 (J \rho_{13} + \gamma_{z} \rho_{14} - J \rho_{15})  + h_{y,2} (-1 + \rho_{1} + \rho_{2} + \rho_{3}) + h_{y,1}(\rho_{10} + \rho_{8})
  - h_{x,1} (\rho_{11} + \rho_{9}),\\
  \partial_{t} \rho_{15} =&   h_{y,1} \rho_{11} + 2 J \rho_{12} - 2 J \rho_{14} - 2 \gamma_{z} \rho_{15} - h_{x,2}  (-1 + \rho_{1} + \rho_{2} + \rho_{3}) + h_{x,1} (\rho_{10} - \rho_{8})
   - h_{y,1} \rho_{9}.
\end{align}
We initialize the state of the system in a product state
\begin{equation}
    \rho(0)=\rho_1\otimes\rho_2,
\end{equation}
of the individual initial qubit states $\rho_i=\frac{1}{2}(1+\rho_x^i\sigma_x^i+\rho_y^i\sigma_y^i+\rho_z^i\sigma_z^i)$, where the vector components are sampled as $\rho_x^i, \rho_y^i, \rho_z^i\sim \mathcal{N}(0,1)$.
We consider the CNOT target transformation
\begin{equation}
    V = \begin{pmatrix}  1 & 0 & 0 & 0 \\ 0 & 1 &  0 & 0\\ 0 & 0 & 0 & 1\\ 0& 0& 1&0 \end{pmatrix},
\end{equation}
and the corresponding fidelity therefore reads
\begin{align}
    F_{\theta}(\vec{\rho}(0), \vec{\rho}(t_f)) =& \Tr(\rho_{f}^{\dagger} V \rho(0) V^{\dagger})\\ 
    =&\rho_{1}(t_f) \rho_{1}(0) + 2 \rho_{6}(t_f) \rho_{10}(0) + 2 \rho_{7}(t_f) \rho_{11}(0) + 2 \rho_{8}(t_f) \rho_{12}(0) + 2 \rho_{9}(t_f) \rho_{13}(0) \label{app-fidelity}\\&
    + 2 \rho_{14}(t_f) \rho_{14}(0) - 2 \rho_{15}(t_f) \rho_{15}(0) +   \rho_{2}(t_f) \rho_{2}(0) + \rho_{3}(0)  \nonumber\\&- 
    (\rho_{1}(t_f) + \rho_{2}(t_f)) \rho_{3}(0)- \rho_{3}(t_f) (-\text{Tr}(\vec{\rho}(t_f))+ \rho_{1}(0) + \rho_{2}(0) + 2 \rho_{3}(0))
   \nonumber\\&+ 
 2 (\rho_{4}(t_f) \rho_{4}(0)
  + \rho_{5}(t_f) \rho_{5}(0)
  + \rho_{10}(t_f) \rho_{6}(0)
  + \rho_{11}(t_f) \rho_{7}(0) + 
    \rho_{12}(t_f) \rho_{8}(0) + \rho_{13}(t_f) \rho_{9}(0)).\nonumber
\end{align}
Instead of evaluating this expression directly, we draw a random time $t_r\in[0,t_f]$ and propagate the initial state to that time which gives us $\rho(t_r)$. We then consider the time-local perturbation proportional to one of the accessible control operators of the Hamiltonian $H_\theta$, i.e. $B_j\in \{\sigma_x^1, \sigma_x^2, \sigma_y^1,\sigma_y^2,\vec{\sigma_1}\vec{\sigma_2}\}$.
In the vector representation $\vec{\rho}$, the corresponding perturbations evaluate as
\begin{align}
    i[\sigma_x^1,{\rho}]\rightarrow&(-2\rho_{7}, -2\rho_{13}, 2\rho_{7}, -\rho_{11} - \rho_{9}, \rho_{10} - \rho_{8}, 0, \rho_{1} - \rho_{3}, -\rho_{15} + \rho_{5}, \\&-\rho_{14} + \rho_{4}, \rho_{15} - \rho_{5}, -\rho_{14} + \rho_{4}, 0, \rho_{1} + 2\rho_{2} + \rho_{3} - 1, \rho_{11} + \rho_{9}, -\rho_{10} + \rho_{8})^T,\nonumber\\
    i[\sigma_x^2,{\rho}]\rightarrow&(-2\rho_{5}, 2\rho_{5}, -2\rho_{15}, 0, \rho_{1} - \rho_{2}, -\rho_{11} + \rho_{9}, \rho_{10} - \rho_{8}, -\rho_{13} + \rho_{7}, \rho_{12} \\&- \rho_{6}, \rho_{13} - \rho_{7}, -\rho_{12} + \rho_{6}, \rho_{11} - \rho_{9}, -\rho_{10} + \rho_{8}, 0, \rho_{1} + \rho_{2} + 2\rho_{3} - 1)^T,\nonumber\\
    i[\sigma_y^1,{\rho}]\rightarrow&(2\rho_{6}, 2\rho_{12}, -2\rho_{6}, \rho_{10} + \rho_{8}, \rho_{11} - \rho_{9}, -\rho_{1} + \rho_{3}, 0, \rho_{14} - \rho_{4}, -\rho_{15} + \\&\rho_{5}, \rho_{14} - \rho_{4}, \rho_{15} - \rho_{5}, -\rho_{1} - 2\rho_{2} - \rho_{3} + 1, 0, -\rho_{10} - \rho_{8}, -\rho_{11} + \rho_{9})^T,\nonumber\\
    i[\sigma_y^2,{\rho}]\rightarrow&(2\rho_{4}, -2\rho_{4}, 2\rho_{14}, -\rho_{1} + \rho_{2}, 0, \rho_{10} + \rho_{8}, \rho_{11} + \rho_{9}, \rho_{12} - \rho_{6}, \rho_{13} - \\&\rho_{7}, \rho_{12} - \rho_{6}, \rho_{13} - \rho_{7}, -\rho_{10} - \rho_{8}, -\rho_{11} - \rho_{9}, -\rho_{1} - \rho_{2} - 2\rho_{3} + 1, 0)^T,\nonumber\\
    i[\vec{\sigma_1}\vec{\sigma_2},{\rho}]\rightarrow&(0, -4\rho_{9}, 4\rho_{9}, 2\rho_{5} - 2\rho_{7}, -2\rho_{4} + 2\rho_{6}, -2\rho_{5} + 2\rho_{7}, 2\rho_{4} - 2\rho_{6}, 0, 2\rho_{2} - \rho_{3}\\&2, 0, 0, -2\rho_{13} + 2\rho_{15}, 2\rho_{12} - 2\rho_{14}, 2\rho_{13} - 2\rho_{15}, -2\rho_{12} + 2\rho_{14})^T.\nonumber
\end{align}
After propagating these resulting operators from $t_r$ to $t_f$, we evaluate the susceptibility of the fidelity in Eq.~(\ref{app-fidelity}) with respect to the perturbation of $B_j$ at time $t_r$, which is 
\begin{eqnarray}
\chi_{j}(t_{r}) = \frac{i}{\hbar} \Tr\Big( V \rho(0) V^{\dagger}  U_\theta(t_{r}, t_{f}) \Big[  B_{j},  U_\theta(t_{r}) \rho(0) U_\theta^{\dagger}(t_{r})\Big]  U_\theta^{\dagger}(t_{r}, t_{f}) \Big).
\end{eqnarray}
We finally update the parameters according to PEPR as described in Eq.~(\ref{eq:theta-update-lr-sine}) in the main-text. It is
\begin{eqnarray}
\theta_{j,k} &\rightarrow & \theta_{j,k} - \alpha \sin(\pi k \frac{t_{r}}{t_f}) \chi_{j}(t_{r}).
\end{eqnarray}
Note, that this update requires only a single time evolution to obtain $\chi_j$, which is used to update $n_j$-many parameters $\theta_{j,k}$. This provides better scaling with respect to the number of parameters, compared to GRAPE.

\begin{figure}
    \centering
    \includegraphics{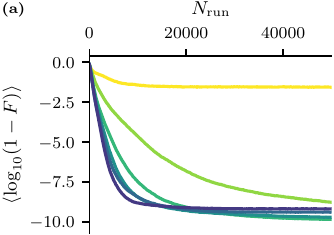}
    \includegraphics{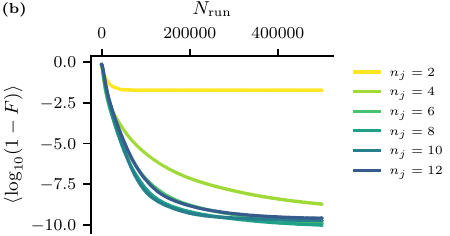}
    \caption{\textbf{Number of Modes.} 
        The average infidelity $\langle \text{log}_{10}(1-F)\rangle$, as a function of the number of runs $N_\text{run}$, for different values of the number of modes $n_{j}$ per control function.
        The lower bound of the infidelity is determined by the time discretization $h$ of the fourth-order Runge-Kutta method.
        Panel \textbf{(a)} shows the results for PEPR with a learning rate of $\alpha_\text{PEPR}=0.3$.
        Panel \textbf{(b)} shows the results for GRAPE with a learning rate of $\alpha_\text{GRAPE}=0.2$ and $\epsilon=10^{-6}$.}
    \label{fig:n_j}
\end{figure}

The expressibility of the control functions is determined by the number of modes $n_j$.
In Fig.~\ref{fig:n_j} we show the average infidelity $\langle \text{log}_{10}(1-F)\rangle$, as a function of the number of runs $N_\text{run}$ based on PEPR and GRAPE for different values of $n_{j}$.
For a value of $n_j=2$, the average infidelity obtains high values, indicating insufficient expressibility of the control functions.
For larger values of $n_j > 2$, the average infidelity converges to a lower bound determined by numerical accuracy.
We find that a value of $n_j=8$ leads to an efficient convergence behavior of the average infidelity, for both PEPR and GRAPE.

\begin{figure}
    \centering
    \includegraphics{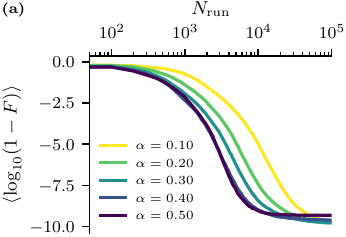}
    \includegraphics{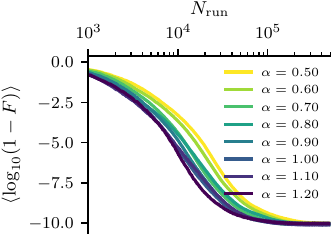}
    \includegraphics{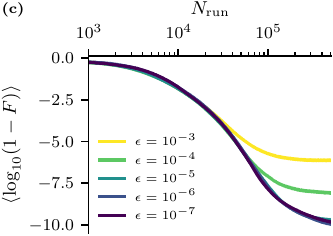}
    \caption{\textbf{Hyperparameters}.
        The average infidelity $\langle \log_{10}(1-F)\rangle$, as a function of the number of runs $N_\text{run}$ for different values of the hyperparameters with $n_{j}=8$.
        The lower bound of the infidelity is determined by the time discretization $h$ of the fourth-order Runge-Kutta method.
        Panel \textbf{(a)} shows the average infidelity based on PEPR for different values of the learning rate $\alpha$.
        Panel \textbf{(b)} shows the average infidelity based on GRAPE for different values of the learning rate $\alpha$.
        Panel \textbf{(c)} shows the average infidelity based on GRAPE for different values of the finite difference length $\epsilon$.}
    \label{fig:hyperparameters}
\end{figure}

We determine an optimal set of hyperparameters of $\alpha_\text{PEPR}=0.5$ for PEPR, and $\alpha_\text{GRAPE}=1.2$, and $\epsilon=10^{-7}$ for GRAPE by comparing the convergence behavior of the average infidelity Eq.~(\ref{eq:infidelity-average}) for different values of the hyperparameters.
In Fig.~\ref{fig:hyperparameters} we show the results that motivate this particular choice of hyperparameters.

\section{Variance of the infidelity}
\label{app:var}

\begin{figure}
    \centering
    \includegraphics{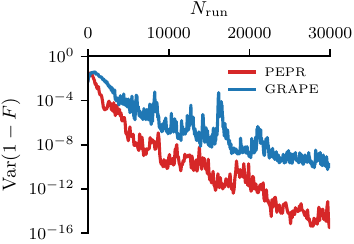}
    \caption{\textbf{Variance of the infidelity.} 
        The variance of the infidelity $\text{Var}(1-F)$ obtained using PEPR (red line) and GRAPE (blue line), as a function of the number of runs $N_\text{run}$.
        PEPR leads to a reduced variance, compared to GRAPE. }
    \label{fig:variance}
\end{figure}

In this section we show the variance of the ensemble of optimization trajectories obtained using PEPR and GRAPE.
The unbiased sample variance over the ensemble of realizations is
\begin{eqnarray}
    \text{Var}(1-F) = \frac{n_T}{n_T-1}\sum_\theta (F_\theta-\langle F_\theta \rangle)^2
\end{eqnarray}
In Fig.~\ref{fig:variance}, we show the variance of the ensembles of realizations for the underlying data used in Fig.~\ref{fig:2}.
PEPR leads to a reduced variance, compared to GRAPE.

\section{Hadamard Gate}
\begin{figure}
    \centering
    \includegraphics{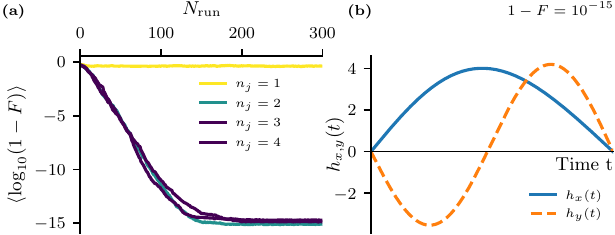}
    \caption{\textbf{Results of PEPR for the Hadamard gate}. 
        Panel \textbf{(a)} shows the average infidelity $\langle \log_{10}(1-F)\rangle$, as a function of the number of runs $N_\text{run}$, for different values of $n_{j}=1,2,3,4$ and for a fixed learning rate of value $\alpha=2.5$.
        The lower bound of the infidelity is determined by the time discretization $h$ of the fourth-order Runge-Kutta method.
        Panel \textbf{(b)} shows the control functions $h_{x}(t)$ and $h_y(t)$ of an example for a high-fidelity implementation for $n_{j}=2$.}
    \label{fig:PEPR-hadamard}
\end{figure}
Additionally to the optimization of the CNOT gate we demonstrate PEPR here for the example of the Hadamard gate on a single qubit.
For this, we consider the Hamiltonian
\begin{equation}
    H_\theta(t) = h_x(t) \sigma_x + h_y(t) \sigma_y,
    \label{eq:hamilton-single-qubit}
\end{equation}
with
$h_x(t)=\sum_k \theta_{x,k}\sin(\pi k t)$ and $h_y(t)=\sum_k \theta_{y,k}\sin(\pi k t)$.
We write the  density operator of the system as
\begin{equation}
    \rho = \frac{1}{2}\begin{pmatrix}
        \xi+\rho_z&\rho_x-i\rho_y\\\rho_x+i\rho_y&\xi-\rho_z
    \end{pmatrix},
    \label{eq:hadamard-rho}
\end{equation}
where $\rho_i\in\mathbb{R}$, and $\xi\in\{0,1\}$ defines the trace of the operator.
We note that $\xi=0$ is required to capture commutator-objects such as $i[B_j,\rho]$ after the perturbation with the control operator $B_j$, as $\text{Tr}([B_j,\rho])=0$. For numerical purposes, we represent $\rho$ as the real-valued vector 
\begin{equation}
    \vec{\rho} = (\rho_x, \rho_y, \rho_z)^T
\end{equation}
The dynamics of the state obey the von-Neumann equation $i\hbar\dot{{\rho}}=[H,{\rho}]$ and the equations of motion read
\begin{align}
    \dot{\rho}_x &= 2h_y\rho_z,\\
    \dot{\rho}_y &= -2h_x\rho_z,\\
    \dot{\rho}_z &= 2h_x\rho_y-2h_y\rho_x.
\end{align}
We initialize the state of the system as $\vec{\rho}(0)\sim \mathcal{N}^3(0,1)$.
We denote the time-propagated state over $[0,t_f]$ as $\rho(t_f)$.
We consider the example of the Hadamard target transformation
\begin{equation}
    V = \frac{1}{\sqrt{2}}\begin{pmatrix}
        1&1\\1&-1
    \end{pmatrix}.
\end{equation}
The fidelity to reach the target state $V\rho(0)V^\dagger$, given the initial state $\rho(0)$, is
\begin{align}
    F_\theta &= \text{Tr}(V\rho(t_0)V^\dagger\rho(t_f)) 
    = \frac{1}{2}(\xi+\rho_z(t_f)\rho_x(t_0)+\rho_x(t_f)\rho_z(t_0)-\rho_y(t_f)\rho_y(t_0)).
\end{align}
The control operators are $B_j\in\{\sigma_x, \sigma_y\}$. In the vector representation $\vec{\rho}$, the corresponding perturbations evaluate as
\begin{align}
    i[\sigma_x, \rho] \rightarrow (0,\rho_z, -\rho_y)^T, \\
    i[\sigma_y, \rho] \rightarrow (-\rho_z, 0, \rho_x)^T.
\end{align}
We update the parameters according to PEPR as described in Eq.~(\ref{eq:theta-update-lr-sine}). 
It is
\begin{eqnarray}
\theta_{j,k} &\rightarrow & \theta_{j,k} - \alpha \sin(\pi k \frac{t_{r}}{t_f}) \chi_{j}(t_{r})
\end{eqnarray}
with the susceptibility under a randomly chosen perturbation $B_j$ at the random time $t_r\in[0,1]$
\begin{eqnarray}
\chi_{j}(t_{r}) = \frac{i}{\hbar} \Tr\Big( V \rho(0) V^{\dagger}  U_\theta(t_{r}, t_{f}) \Big[  B_{j},  U_\theta(t_{r}) \rho(0) U_\theta^{\dagger}(t_{r})\Big]  U_\theta^{\dagger}(t_{r}, t_{f}) \Big).
\end{eqnarray}
$U_\theta$ denotes the unitary time-evolution operator generated by the Hamiltonian Eq.~(\ref{eq:hamilton-single-qubit}).

In Fig.~\ref{fig:PEPR-hadamard}~(a) we show the average infidelity $\langle \log_{10}(1-F)\rangle$, as a function of the number of runs $N_\text{run}$ for an empirically determined optimal learning rate of $\alpha_\text{PEPR}=2.5$.
The minimal number of modes to obtain high-fidelity protocols in this simple example is $n_{j} > 1$.
The average infidelity converges to values of $\langle \log_{10}(1-F)\rangle(N_\text{run})\rightarrow 10^{-15}$, for $N_\text{run} > 200$.
The lower bound of the infidelity is determined by the time discretization $h$ of the fourth-order Runge-Kutta method.
We show an example for a high-fidelity implementation of the Hadamard gate with $n_{j}=2$ in Fig.~\ref{fig:PEPR-hadamard}~(b).

\end{document}